\documentclass{aastex631}

\let\tablenum\relax
\usepackage{siunitx}
\DeclareSIUnit\year{yr}
\DeclareSIUnit\day{day}
\usepackage[version=4]{mhchem}

\definecolor{C0}{HTML}{e24a33}
\definecolor{C1}{HTML}{348abd}
\definecolor{C2}{HTML}{988ed5}
\definecolor{C3}{HTML}{777777}
\usepackage{xspace}
\newcommand{\expSE}{\textcolor{black}{\emph{StretchExplicit}}\xspace}
\newcommand{\expSR}{\textcolor{black}{\emph{StretchReduced}}\xspace}
\newcommand{\expSP}{\textcolor{black}{\emph{StretchParam}}\xspace}
\newcommand{\expUP}{\textcolor{black}{\emph{UniformParam}}\xspace}

\shorttitle{Stretched-mesh simulations for a climate of a tidally locked exoplanet}
\shortauthors{Sergeev et al.}

\begin{document}

\title{The impact of the explicit representation of convection on the climate of a tidally locked planet in global stretched-mesh simulations.}

\correspondingauthor{Denis E. Sergeev}
\email{d.sergeev@exeter.ac.uk}

\author[0000-0001-8832-5288]{Denis E. Sergeev}
\affiliation{Department of Physics and Astronomy, University of Exeter, Exeter, EX4 4QL, UK}

\author[0000-0002-1485-4475]{Ian A. Boutle}
\affiliation{Met Office, Fitzroy Road, Exeter, EX1 3PB, UK}
\affiliation{Department of Physics and Astronomy, University of Exeter, Exeter, EX4 4QL, UK}

\author[0000-0002-4664-1327]{F. Hugo Lambert}
\affiliation{Department of Mathematics and Statistics, University of Exeter, Exeter, EX4 4QF, UK}

\author[0000-0001-6707-4563]{Nathan J. Mayne}
\affiliation{Department of Physics and Astronomy, University of Exeter, Exeter, EX4 4QL, UK}

\author[0000-0002-6546-8400]{Thomas Bendall}
\affiliation{Met Office, Fitzroy Road, Exeter, EX1 3PB, UK}

\author[0000-0003-0165-4885]{Krisztian Kohary}
\affiliation{Department of Physics and Astronomy, University of Exeter, Exeter, EX4 4QL, UK}

\author{Enrico Olivier}
\affiliation{Research Software Engineering, University of Exeter, Exeter, EX4 4QE, UK}

\author{Ben Shipway}
\affiliation{Met Office, Fitzroy Road, Exeter, EX1 3PB, UK}

\begin{abstract} 
Convective processes are crucial in shaping exoplanetary atmospheres but are computationally expensive to simulate directly.
A novel technique of simulating moist convection on tidally locked exoplanets is to use a global 3D model with a stretched mesh.
This allows us to locally refine the model resolution to 4.7~km and resolve fine-scale convective processes without relying on parameterizations.
We explore the impact of mesh stretching on the climate of a slowly rotating TRAPPIST-1e-like planet, assuming it is 1:1 tidally locked.
In the stretched-mesh simulation with explicit convection, the climate is 5~K colder and 25\% drier than that in the simulations with parameterized convection (with both stretched and quasi-uniform meshes).
This is due to the increased cloud reflectivity --- because of an increase of low-level cloudiness --- and exacerbated by the diminished greenhouse effect due to less water vapor.
At the same time, our stretched-mesh simulations reproduce the key characteristics of the global climate of tidally locked rocky exoplanets, without any noticeable numerical artifacts.
Our methodology opens an exciting and computationally feasible avenue for improving our understanding of 3D mixing in exoplanetary atmospheres.
Our study also demonstrates the feasibility of a global stretched mesh configuration for LFRic-Atmosphere, the next-generation Met Office climate and weather model.
\end{abstract}


\section{Introduction}
Fine-scale atmospheric phenomena such as moist convection, are typically unresolved in exoplanet climate models because the grid in these models is too coarse.
Therefore, most general circulation models (GCMs) rely on convection parameterizations \citep{Arakawa04_cumulus}.
While convection parameterizations are physically motivated, they are necessarily a simplification of the real process \citep{Kendon21_challenges, Rios-Berrios22_differences}.
Recent studies showed that resolving convection explicitly may impact the estimate of the global climate of terrestrial tidally locked exoplanets \citep{Sergeev20_atmospheric, Lefevre21_3d, Yang23_cloud}, because of the cloud stabilizing feedback \citep{Yang13_stabilizing}.
However, global convection-resolving simulations are computationally challenging, especially for climate-scale runs.

Locally refined global grids offer an elegant solution.
They allow one to focus on a specific area of the planet, while keeping the global coverage and allowing for interactions at different scales, e.g. between localized convection and planetary-scale water distribution.
Additionally, locally refined grids avoid numerical artifacts associated with using limited-area, or regional, models \citep{Fox-Rabinovitz08_stretchedgrid, Uchida16_error}.
Namely, there are no boundaries or sharp changes in resolutions between the high-resolution region and the global model.
These grids also allow for the two-way interaction between the region of interest and the rest of the planet, unlike a typical regional setup as in e.g. \citet{Sergeev20_atmospheric}.
This is especially important for the climate of tidally locked exoplanets, for which the localized stellar forcing is the key driving mechanism \citep[e.g.][]{Wordsworth22_atmospheres}.

A form of localized resolution increase is to stretch or deform the global mesh \citep[e.g.][]{Fox-Rabinovitz00_simulation, Uchida16_error, Harris16_high-resolution, Bindle21_grid-stretching}.
The advantages of this method are that the mesh remains topologically the same and does not require modifications to the model transport schemes.
Grid stretching is also performed gradually, resulting in fewer grid artifacts compared to other methods.
On the other hand, grid stretching typically results in a lower resolution on the opposite side of the planet, though it was not found to noticeably degrade the solution \citep{Harris16_high-resolution}.

Stretched grids have been successfully applied for modeling Earth's atmosphere, bringing improvements in reproducing phenomena such as tropical precipitation \citep{Harris16_high-resolution} and localized greenhouse gas emissions \citep{Bindle21_grid-stretching}.
The topological simplicity of stretched grids has been shown to be particularly useful for simulating extreme precipitation and tropical cyclones in aquaplanet Earth simulations \citep{Harris16_high-resolution}.
Despite a substantial degree of stretching, there was no degradation of the global climate in these simulations and no spurious numerical noise contaminating the solution.
An intercomparison of several stretched-mesh GCMs was shown to successfully reproduce the key aspects of regional climate \citep{Fox-Rabinovitz06_variable}, in which the major positive impact was due to a better representation of model dynamics and orography \citep{Fox-Rabinovitz08_stretchedgrid}.

Tidally locked exoplanets offer an excellent use case for stretched-mesh GCMs because convective processes are hypothesized to occur mostly on the hot day side, on which the local refinement can be centered.
Our work is the first application of a stretched-mesh GCM for the case of a tidally locked terrestrial exoplanet.
We perform four experiments, in which we move from a configuration with a quasi-uniform mesh with parameterized convection to a stretched mesh with explicit convection.
This includes an experiment with a ``reduced'' parameterization, serving as an intermediate step between fully parameterized and explicit convection.
We demonstrate that it is computationally feasible to maintain the simulation fidelity of the global climate and at the same time resolve mesoscale circulations associated with the moist convection.
Our stretched-mesh simulations resolve fine-scale cloud patterns on the day side to a similar degree as was done previously in limited-area \citep{Zhang17_surface, Sergeev20_atmospheric, Lefevre21_3d} and pseudo-global \citep{Yang23_cloud} high-resolution simulations.
Using an explicit convection configuration on a stretched mesh leads to a colder, lower relative humidity climate, and an increase in cloud water content and the rate of the most intense precipitation.
At the same time, the structure of the large-scale atmospheric circulation typical for tidally locked terrestrial exoplanets is preserved, and mesh stretching does not produce any numerical artifacts.

\section{Methods}
\label{sec:methods}
We use LFRic-Atmosphere, the new 3D GCM of the Met Office \citep{Adams19_lfric}, based on a novel dynamical core GungHo and a suite of well-tested physical parameterizations inherited from its forerunner, the Unified Model (UM).
GungHo solves the fully compressible non-hydrostatic Euler equations on a quasi-uniform cubed-sphere mesh using a mimetic finite-element discretisation and a mass-conserving finite-volume transport scheme \citep{Melvin19_mixed, Bendall20_compatible, Bendall22_solution, Kent23_mixed}.
The radiative transfer is parameterized using a two-stream correlated-$k$ scheme SOCRATES based on \citet{Edwards96_studies}, for which we use the same setup as in \citet{Sergeev22_thai}.
LFRic-Atmosphere reproduces a variety of atmospheric flows \citep{Melvin19_mixed, Kent23_mixed, Brown23_physicsdynamicschemistry, Melvin24_mixed}, including climate benchmarks for terrestrial exoplanets \citep{Sergeev23_simulations}.

\begin{figure*}[ht!]
    \centering
    \includegraphics[width=\textwidth]{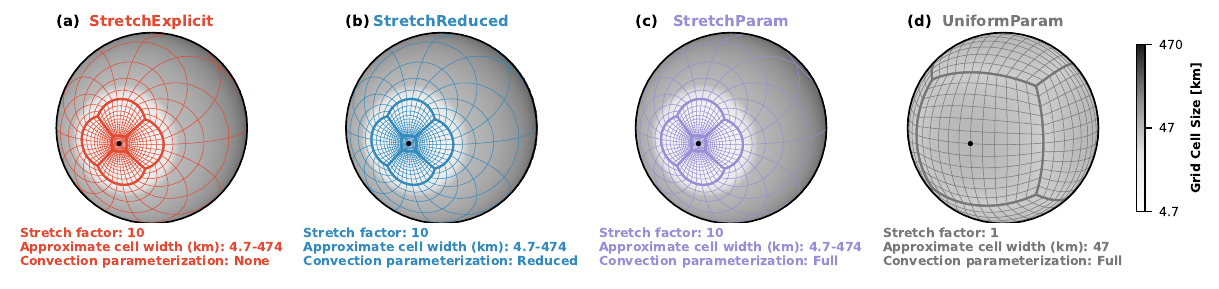}
    \caption{%
    Summary of the simulation setup.
    Grey shading shows the cell width (\si{\km}) calculated as a square root of the cell area.
    All simulations use the C192 mesh.
    Thick lines are the boundaries of cubed-sphere faces, and thin lines are the boundaries of every 16$\times$16 grid cells.
    The black dot shows the location of the substellar point.
    For more details on the stretching procedure, see Appendix~\ref{sec:mesh_stretch}.
    Here and throughout the paper, the experiments are color-coded: \expSE is shown in red, \expSR in blue, \expSP in purple, and \expUP in gray.
    }
\label{fig:mesh_geom}
\end{figure*}


Our simulation parameters are based on the TRAPPIST-1 Habitable Atmosphere Intercomparison \citep[THAI,][]{Fauchez20_thai_protocol}, a set of steady-state climate scenarios for a tidally locked TRAPPIST-1e assuming a 1 bar atmosphere.
\citet{Sergeev23_simulations} have shown that LFRic-Atmosphere reproduces the THAI scenarios close to that obtained from other models \citep[e.g.][]{Turbet22_thai, Wolf22_exocam, Paradise22_exoplasim}.
Here, we use the nitrogen-dominated aquaplanet case, \emph{Hab~1} \citep{Sergeev22_thai}, to highlight the impact of mesh stretching on cloud regimes and moist convection.

Importantly, we set the rotation period of the planet to be 12.2 days, i.e. twice as long as the one used in the THAI scenarios, while still assuming the planet is tidally locked.
This is to avoid the climate bistability that often appears in 3D GCM simulations of the \emph{Hab~1} case \citep{Sergeev22_bistability} and would complicate our analysis, which is focused on the impact of the grid resolution.
However, for any given case of a tidally-locked exoplanet within a similar region of parameter space, our findings are still pertinent. 
Additional simulations with the original rotation rate of 6.1 days indeed result in two distinct climate regimes, as we discuss in Appendix~\ref{sec:bistability}.

Throughout this study we use the C192 cubed-sphere mesh, in a stretched and non-stretched configuration (see Appendix~\ref{sec:mesh_stretch} for more details on mesh stretching).
We performed four experiments: three with the mesh stretched by a factor of 10, and one with a non-stretched, quasi-uniform mesh (Fig~\ref{fig:mesh_geom}), Appendix~\ref{sec:mesh_stretch}).
In the \expUP and \expSP configuration we use the same parameterizations as those used in \citet{Sergeev23_simulations}, including the standard mass-flux parameterization of convection.
The \expSP experiment differs from \expUP only in the mesh configuration, while the \expSR experiment uses a ``reduced'' convective parameterization (for technical details, see Appendix~\ref{sec:rc_setup}).
The \expSR configuration is designed to allow small, unresolved convection to be handled by the parameterization, whilst allowing deeper convection to be explicitly resolved.
Note that the transition between resolved and parameterized convection in \expSR is not defined by a specific threshold.
Instead, it dynamically changes with the cell size (and is therefore particularly useful in a stretched-mesh setup): when the scale of convection is smaller than the grid scale, convection is handled by the parameterization, whilst when the scale of convection is greater than the grid scale, parameterization effectively switches off and convection is handled explicitly.
The \expSR experiment thus serves as an intermediate step between fully parameterized and fully explicit convection.
The \expSE configuration handles convection explicitly, i.e. with the parameterization switched off \citep[see discussion in][]{Sergeev20_atmospheric}.

All four experiments have 63 levels with a model top at about \SI{41}{\km}.
This corresponds to the same vertical resolution below \SI{41}{\km} as that used by the Met Office's Unified Model in global climate simulations \citep{Walters19_ga7, Tomassini23_confronting}.
While the vertical resolution may impact the degree of convective aggregation \citep{Jenney23_vertical}, the simulations for tidally locked terrestrial planets have not been sensitive to it \citep{Wei20_small, Yang23_cloud}.
The impact of the vertical resolution in a stretched-mesh configuration will be investigated in a future study. 

All simulations start from a dry isothermal (\SI{300}{\K}) atmosphere at rest \citep{Fauchez20_thai_protocol}.
We integrate the model for 1000 Earth days, which is sufficient, given the relatively thin atmosphere and the shallow (\SI{1}{\m}) slab ocean at the bottom boundary (judging by the surface temperature and top-of-atmosphere energy balance reaching a steady state). 
Discarding the spin-up period, in Sec.~\ref{sec:results} we present climate diagnostics for the final 500 days.

\section{Results}
\label{sec:results}
\begin{figure*}
    \centering
    \includegraphics[width=1\linewidth]{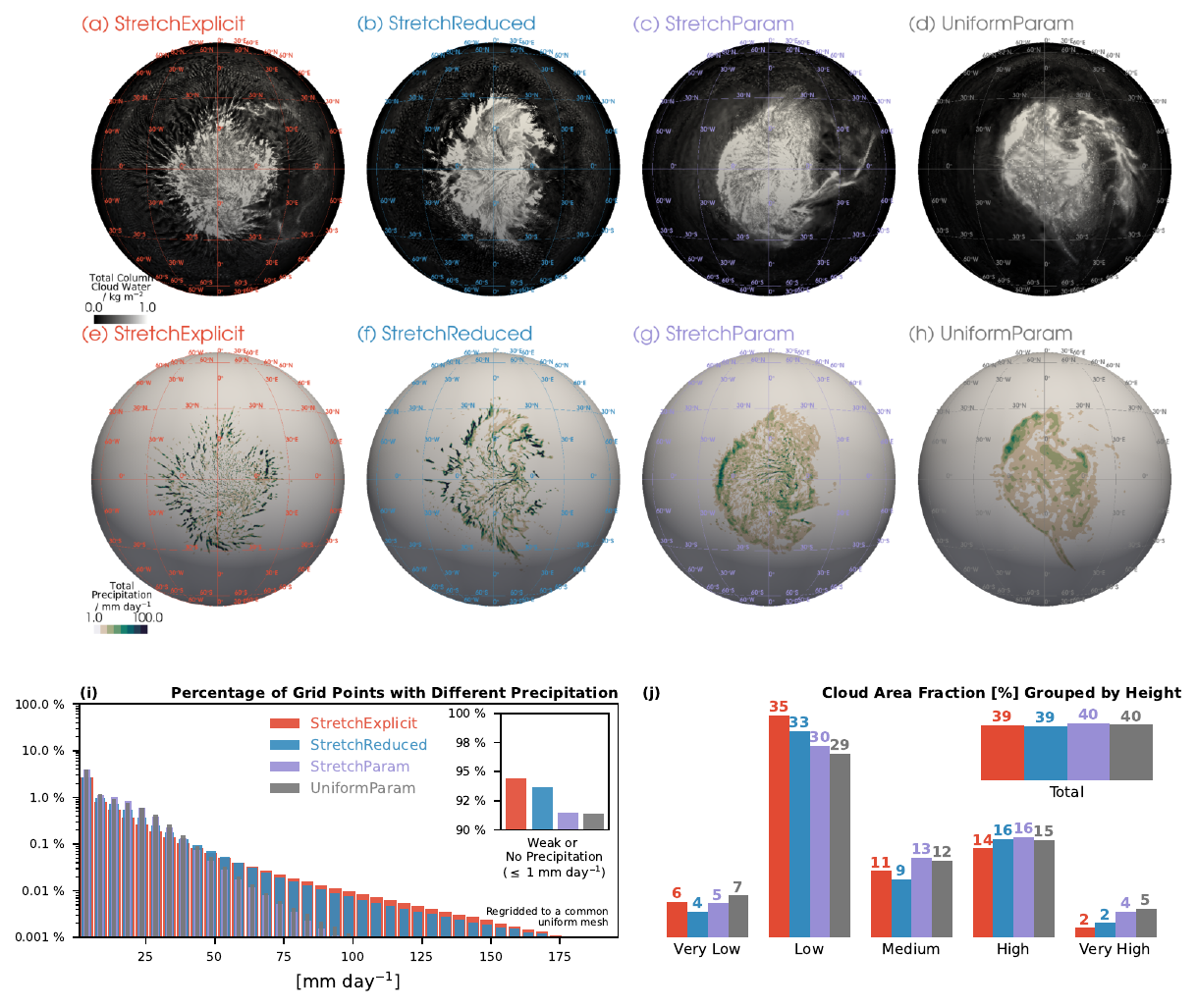}
    \caption{%
    Clouds and precipitation in the simulations with stretched (\expSE, \expSR, \expSP) and quasi-uniform (\expUP) mesh.
    \textbf{(a--d)} Maps of instantaneous vertically integrated cloud water (liquid plus ice) at the end of the simulations (\si{\kg\m\squared}).
    \textbf{(e--h)} Maps of instantaneous precipitation rate (\si{\mm\per\day}).
    The substellar point is at the center of the maps (\ang{0}N, \ang{0}E).
    \textbf{(i)} Histogram of instantaneous precipitation rate interpolated to a common quasi-uniform mesh (same mesh as in the \expUP simulation).
    The histogram shows the percentage of grid cells with precipitation above \SI{1}{\mm\per\day}, binned with a step of \SI{5}{\mm\per\day}.
    Note the logarithmic scale of the vertical axis.
    The inset shows the percentage of grid cells with precipitation below \SI{1}{\mm\per\day} (weak or non-precipitating points).
    \textbf{(j)} Bar chart of cloud area fraction (\si{\percent}) grouped by different heights.
    Experimental setup is detailed in Fig~\ref{fig:mesh_geom} and Sec.~\ref{sec:methods}.
    }
    \label{fig:snap_hist}
\end{figure*}

Adopting a refined mesh allows us to simulate convective clouds covering the substellar area in great detail.
As the snapshots in Fig.~\ref{fig:snap_hist}a--d show, the stretched-mesh simulations produce spatial patterns of total column cloud condensate and precipitation rate similar to those reported in a convection-permitting limited-area model \citep{Sergeev20_atmospheric}.
The coarse, quasi-uniform mesh, on the other hand, produces a smoother pattern with a smaller variation of cloudiness and less extreme rainfall.

On the periphery of the substellar region, the stretched-mesh simulation without a convection parameterization (\expSE), produces cloud bands reminiscent of ``cloud streets'' (Fig.~\ref{fig:snap_hist}a).
These parallel elongated bands of clouds with cloud-free conditions between them were reported by \citet{Yang23_cloud} in quasi-global convection-permitting simulations for TRAPPIST-1e.
They form as a result of the persistent advection of cold air from the night side onto a warmer ocean surface of the day side, resulting in a formation of convective boundary layers.
This is similar to the roll convection occurring in cold-air outbreaks on Earth \citep[e.g.][]{Gryschka05_roll}.
Cloud streets can have a cooling effect on the global climate because the outgoing longwave radiation is larger in the clear-sky regions than that in the cloudy regions \citep{Yang23_cloud}.
Our work supports the results from \citet{Yang23_cloud} and suggests that cloud streets can be reproduced in a stretched-mesh global model at a reduced computational cost.
However, they may also be an artifact of under-resolving turbulence because of the increasing grid spacing and the 1D boundary layer parameterization in our setup \citep[for the relevant discussion, see][]{Boutle14_seamless}.

The small-scale variability enabled by the mesh refinement and explicit convection is also reflected in a marked increase in the precipitation rate in the substellar region.
As Fig.~\ref{fig:snap_hist}h shows, the maximum instantaneous precipitation reaches \SI{100}{\mm\per\day} in the \expUP simulation.
Moving from a quasi-uniform to a stretched mesh has little effect (Fig.~\ref{fig:snap_hist}g), while reducing or disabling the convection scheme results in almost doubling of the maximum precipitation rate (Fig.~\ref{fig:snap_hist}e,f).
The histogram in Fig.~\ref{fig:snap_hist}i distills this change in the precipitation maxima.
It shows that the mesh stretching barely changes precipitation rates (after interpolating the data to the same mesh). 
However, reducing the impact of the convection scheme or even entirely switching it off results in substantially more intense precipitation, reaching \SI{175}{\mm\per\day} at least twice within our analysis period of 500 days (\SI{0.001}{\percent} of 221184 mesh cells).
This is because in the \expSE and \expSR cases much of the convection is handled by the grid-scale dynamics, permitting the formation of small but concentrated storms.
Similarly high precipitation rate was reported for the stretched-mesh simulations of Earth as an aquaplanet \citep{Harris16_high-resolution} and for the convection-permitting limited-area simulations of the climate of TRAPPIST-1e and Proxima Centauri b \citep{Sergeev20_atmospheric}.

\begin{figure*}
    \centering
    \includegraphics[width=1\linewidth]{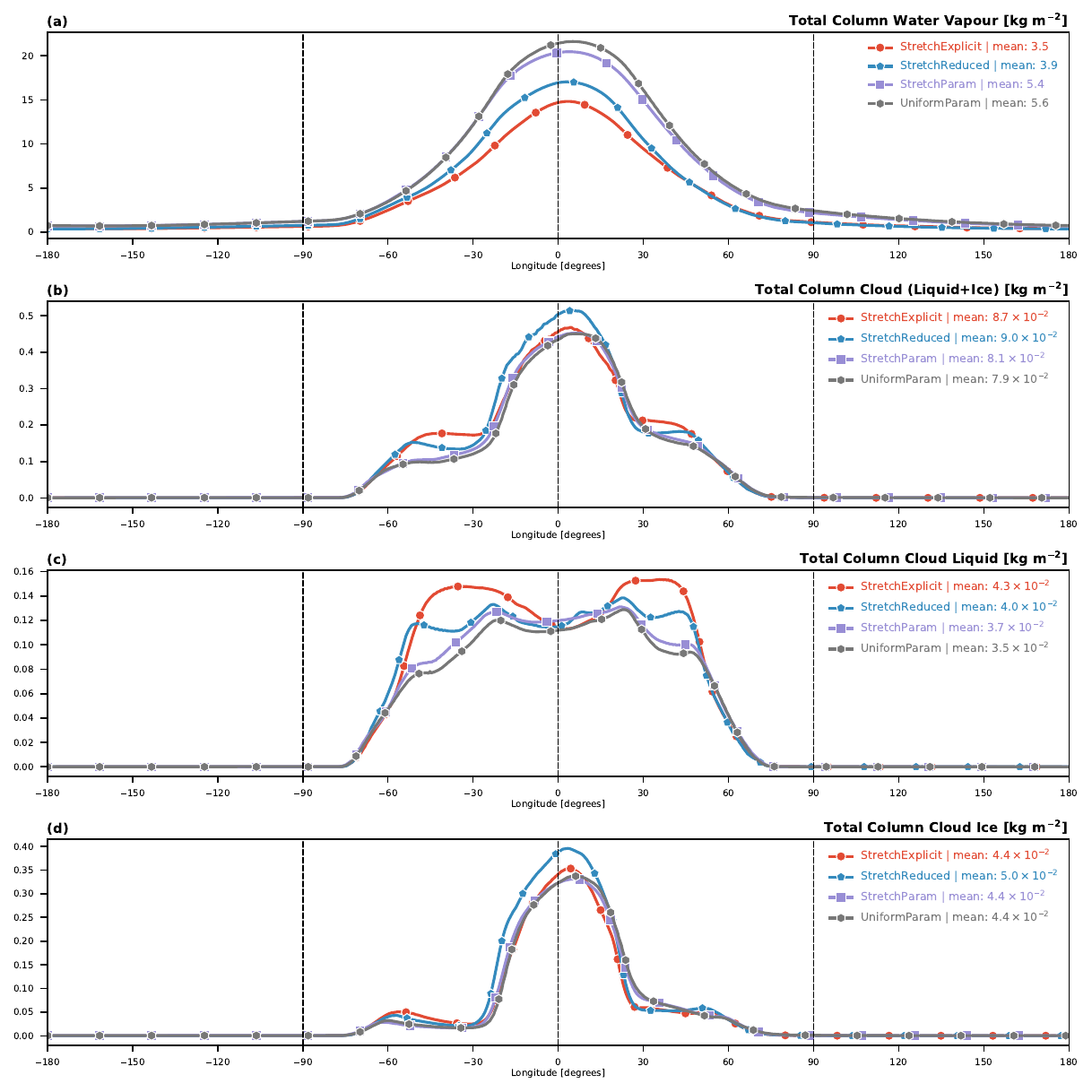}
    \caption{
    Meridional and time mean profiles of vertically integrated moisture diagnostics in \si{\kg\per\m\squared}: (a) water vapor, (b) total cloud (liquid plus ice), (c) cloud liquid, (d) cloud ice.
    }
    \label{fig:cloud_prof}
\end{figure*}

\begin{figure*}
    \centering
    \includegraphics[width=1\linewidth]{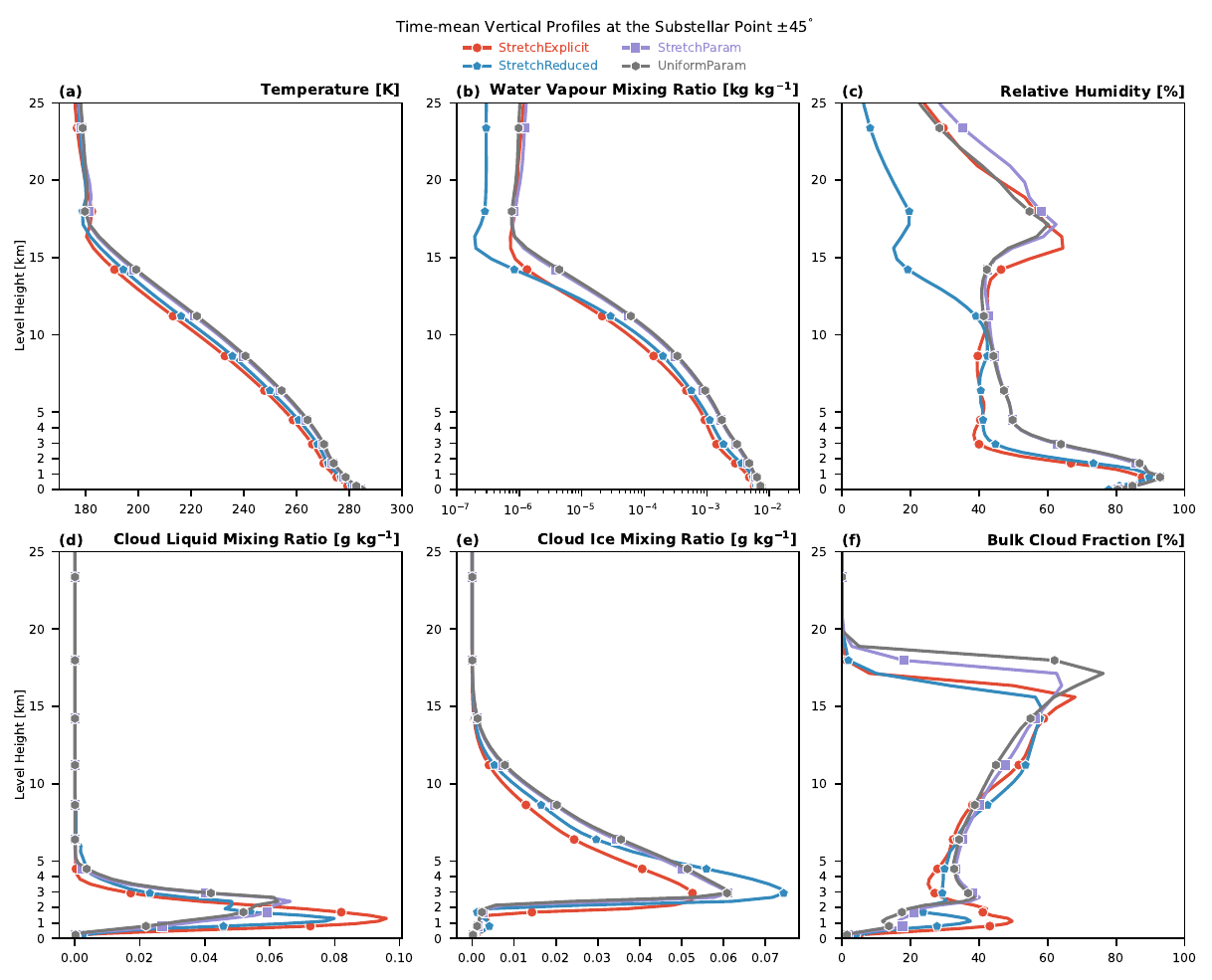}
    \caption{
    Vertical profiles of time mean diagnostics in the substellar region within the bottom \SI{25}{\km} of the atmosphere: \textbf{(a)} temperature in \si{\K}, (b) water vapor mixing ratio in \si{\kg\per\kg}, (c) relative humidity in \si{\percent}, (d) cloud liquid mixing ratio in \si{\g\per\kg}, (e) cloud ice mixing ratio in \si{\g\per\kg}, (f) cloud fraction in \si{\percent}.
    }
    \label{fig:ss_prof}
\end{figure*}

The main effects of explicit convection on the global climate are a reduction in atmospheric water vapor content and an increase in condensed cloud water.
While the \expUP and \expSP cases have a similar amount of water vapor (\SIrange{5.4}{5.6}{\kg\per\m\squared}), in the \expSR and \expSE cases it decreases by a about \SI{25}{\percent}, with the latter simulation being the driest overall (Fig.~\ref{fig:cloud_prof}a).
The overall drying of the atmosphere in the \expSE and \expSR simulations is evident in the vertical profiles of absolute humidity (Fig.~\ref{fig:ss_prof}b) and relative humidity (Fig.~\ref{fig:ss_prof}c).
The relative humidity is noticeably lower in the troposphere in the \expSE and \expSR simulations, so the water vapor decrease cannot be explained only by the lower temperatures in these two simulations (Fig.~\ref{fig:ss_prof}a).
Indeed, since the stratospheric temperature is broadly the same in all four cases, it is the parameterization of turbulent mixing above the boundary layer included in the \expSR configuration that reduces the mixing across the tropopause and results in a less efficient upward transport of water vapor.
This leads to a lower humidity in the upper layers of the night side and terminator regions (not shown), which may have a small effect on the water features in the transmission spectrum as discussed e.g. in \citet{Fauchez22_thai} and \citet{Sergeev22_bistability}.

At the same time, the global amount of cloud water increases from \num{7.9e-2} in the \expUP experiment to \num{9e-2} and \SI{8.7e-2}{\kg\per\m\squared} in the \expSR and \expSE experiments, respectively (Fig.~\ref{fig:cloud_prof}b).
The largest increase of cloud condensate for the \expSE case is seen in the liquid phase (Fig.~\ref{fig:cloud_prof}c).
It happens predominantly in the ring around the substellar region, i.e. where the grid resolution becomes relatively coarse due to the nature of the mesh stretching.
The grid spacing is too large to fully resolve the shallow convection, and the representation of convective updrafts suffers from the lack of parameterization that can remove the instability.
Explicit convection overcompensates for this, and as a result, convective updrafts remove moisture out of the boundary layer too efficiently, creating more cloud condensate (evident especially in the liquid cloud content in Fig.~\ref{fig:cloud_prof}c).
In the \expSR experiment, the parameterizations are set up to represent shallow convection better.
As a result, grid-aliased convection is not as intense, and the cloud content around the substellar region is lower than that in the \expSE case (Fig.~\ref{fig:cloud_prof}c,d).
The liquid cloud increase in the reduced and explicit convection simulations contributes to the changes in the cloud fraction.
As Fig.~\ref{fig:snap_hist}j reveals, moving from parameterized convection (\expSP and \expUP) to reduced/explicit convection (\expSR and \expSE), the global-mean low cloud fraction increases by up to \SI{6}{\percent}.
This change happens predominantly in the substellar region, and is indeed due to the liquid phase increase, as evident in the vertical profiles of the mixing ratio and bulk cloud fraction (Fig.~\ref{fig:ss_prof}d,f).

Compared to previous high-resolution studies of e.g. \citet{Sergeev20_atmospheric}, our results broadly agree that the high-resolution model with explicit convection produces more low-level clouds than a coarse-resolution GCM with parameterized convection.
This is unsurprising because most subgrid-scale physics parameterizations in LFRic-Atmosphere are inherited from the Unified Model \citep{Sergeev23_simulations}.
The differences between our study and that of \citet{Lefevre21_3d} are greater mostly because of the inter-model differences in physical parameterizations.
Moreover, the high-resolution simulation in \citet{Lefevre21_3d} is forced by a constant and spatially uniform heating rate, while in the present study, the impact of the global circulation on the substellar region is direct due to the gradual stretching of the mesh and no boundaries between nested and driving models.
Finally, the planetary parameters are different between the aforementioned studies and the present study.
As a result, the substellar atmosphere is warmer in the simulations in \citet{Lefevre21_3d} and \citet{Sergeev20_atmospheric} than in the present study.
This likely leads to more water being in the vapor form than in the condensed cloud form in those two studies.
Another consequence of a colder climate in our simulations is that the substellar convection is likely to be slightly weaker, resulting in a weaker transport of cloud condensate in the upper atmosphere and hence fewer high clouds.

\begin{figure*}
    \centering
    \includegraphics[width=1\linewidth]{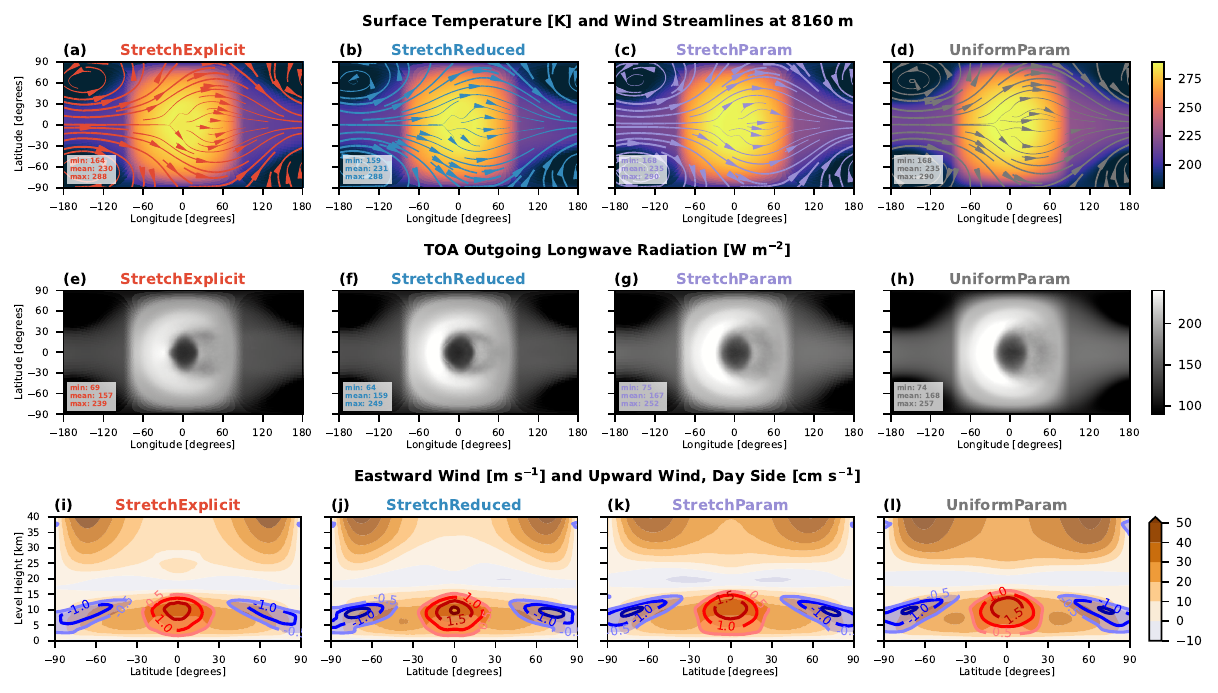}
    \caption{
    Time mean thermodynamic and circulation regime in our simulations.
    \textbf{(a--d)} Surface temperature in \si{\K} overlaid by streamlines of the horizontal wind at \SI{\approx 8}{\km} above the surface.
    \textbf{(e--h)} Top-of-atmosphere outgoing longwave radiation in \si{\watt\per\m\squared}.
    \textbf{(i--l)} Zonal mean of the eastward wind (orange contours, \si{\m\per\s}) overlaid by the day-side upward wind (red and blue contours, \si{\cm\per\s}).
    }
    \label{fig:tsfc_olr_winds}
\end{figure*}

The cloud cover differences between the parameterized and explicit convection experiments lead to changes in the radiation balance and, consequently, the surface temperature.
In the \expSE case, the top-of-atmosphere shortwave cloud radiative effect is \SI{-58.8}{\watt\per\m\squared}, more than \SI{10}{\watt\per\m\squared} greater in magnitude than that in the \expUP case.
A similar brightening of clouds was reported in \citet{Tomassini23_confronting} for the reduced convection experiments.
Expressed in terms of the global mean albedo, the amount of reflected shortwave radiation ranges from \SI{21.9}{\percent} in the \expUP case to \SI{25.2}{\percent} in the \expSE case.
Together with the reduction in the greenhouse effect due to less water vapor, this leads to a \SI{5}{\K} drop in the global mean surface temperature in the \expSE case (Fig.~\ref{fig:tsfc_olr_winds}a--d).
The global mean surface temperature is 230, 231, 235 and \SI{235}{\K} in the \expSE, \expSR, \expSP, and \expUP, respectively.
Note that the \expSR climate is not colder than \expSE despite having a substantially lower stratospheric humidity (Fig.~\ref{fig:ss_prof}b).
This is because the greenhouse effect of the stratospheric water vapor is small due to its very low absolute amount.

The thermal maps in our simulations exhibits a day-night dichotomy typical for tidally locked planets, both in terms of surface temperature (Fig.~\ref{fig:tsfc_olr_winds}a--d) and top-of-atmosphere outgoing longwave radiation (Fig.~\ref{fig:tsfc_olr_winds}e--h).
The substellar area reaches temperatures of \SIrange{288}{290}{\K}, while the night-side coldest areas (stationary gyres) reach temperatures of \SIrange{159}{168}{\K}.
The latter substantially lower than the value predicted by a coarse-mesh LFRic-Atmosphere simulations of the THAI Hab~1 case \citep{Sergeev23_simulations}, which is likely due to the slower rotation rate of the planet in the present study.
A similar effect of slowing the rotation rate on the climate was found in GCM simulations by e.g. \citet{Edson11_atmospheric} and \citet{Komacek19_atmospheric}.
Likewise, the outgoing longwave radiation is at the lower end of the THAI inter-model spread \citep{Sergeev22_thai}: its values range from \SI{167.8}{\watt\per\m\squared} in the \expUP experiment to \SI{156.8}{\watt\per\m\squared} in \expSE (Fig.~\ref{fig:tsfc_olr_winds}e--h).
Clouds prevent even more heat being lost to space, so their net effect on the top-of-atmosphere longwave radiation flux is positive.
It is broadly the same in our four experiments (\SI{\approx 13}{\watt\per\m\squared}) and is comparable to that predicted by the models in \citet{Sergeev22_thai} and \citet{Yang23_cloud}.

In agreement with the global temperature distribution, the large-scale circulation is broadly the same in all four simulations (see streamlines in Fig.~\ref{fig:tsfc_olr_winds}a--d).
The main pattern of the wind field in the troposphere is a superrotating (prograde) jet at the equator and stationary cyclonic gyres in high latitudes on the night side, as is typically expected for synchronously rotating terrestrial planets \citep[e.g.][]{Haqq-Misra18_demarcating, Komacek19_atmospheric, Hammond20_equatorial}.
The vertical structure of the wind field reveals that the atmosphere has two regions of eastward flow: tropospheric (below \SI{\approx 15}{\km}) and stratospheric (above \SI{\approx 25}{\km}), with a quiescent region in between (Fig.~\ref{fig:tsfc_olr_winds}i--l).
The tropospheric flow is characterized by intense upward motions in the substellar region (red contours in Fig.~\ref{fig:tsfc_olr_winds}i--l), associated with convergence near the surface and divergence aloft.
The zonal-mean vertical velocity is marginally stronger in the \expSR simulation, which is due to the changes in the convection parameterization.
Mesh stretching alone brings a relatively small change in the time-averaged vertical velocity (Fig.~\ref{fig:tsfc_olr_winds}k).
In the stratosphere, the eastward flow has two maxima in high latitudes, which become somewhat weaker moving from \expUP to \expSE.
Hence, the 3D wind field in our simulations is broadly similar to and only slightly weaker than that in the Unified Model output for THAI Hab~1 \citep{Sergeev22_thai}.

\begin{figure*}
    \centering
    \includegraphics[width=1\linewidth]{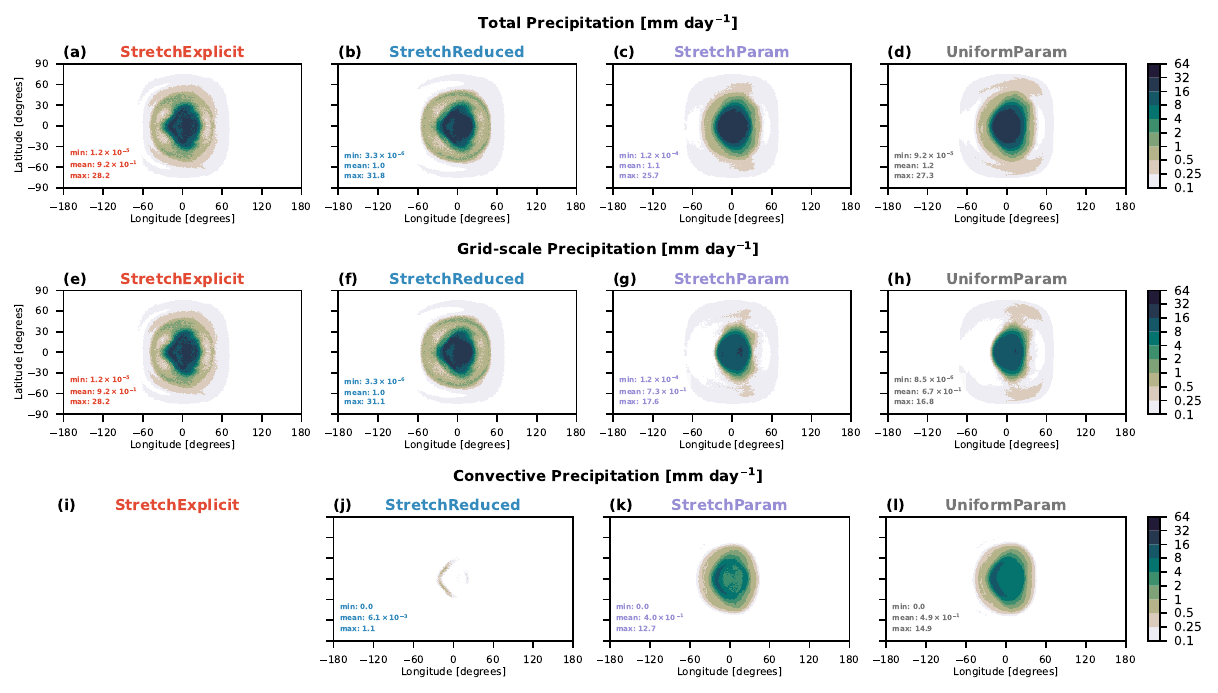}
    \caption{
    Maps of time mean precipitation rate in \si{\mm\per\day}, from top to bottom: total, grid-scale, convective.
    Note that there is no convective precipitation in the \expSE simulation, because the convection parameterization is switched off.
    }
    \label{fig:prec_maps}
\end{figure*}

Despite better resolving small-scale convective patterns (Fig.~\ref{fig:snap_hist}), on average our stretched-mesh simulations produce a similar total precipitation rate as compared to that in the non-stretched-mesh simulations with parameterized convection (Fig.~\ref{fig:prec_maps}).
The precipitation rate has a broad peak over the substellar region, reaching \SI{\approx 30}{\mm\day} at the substellar point --- close to the highest values within the inter-tropical convergence zone in \citet{Harris16_high-resolution}.
Averaged meridionally (across all latitudes), the peak of precipitation is about \SI{10}{\mm\per\day} (not shown) and agrees well with the values in \citet{Yang23_cloud}.
However, the way the model generates precipitation is different.
Grid-scale precipitation, generated by the microphysics scheme, is the largest in the \expSE and \expSR cases, accounting for almost all of the precipitation in these cases (Fig.~\ref{fig:prec_maps}e,f).
Convective precipitation, on the other hand, the largest in the \expUP case, and second-largest in \expSP: \SI{\approx 0.4}{\mm\per\day}  (Fig.~\ref{fig:prec_maps}k,l).
In the \expSR simulation, the reduced convection parameterization yields about two orders of magnitude less precipitation than the full parameterization.
This is because the \expSR simulation has a larger convective available potential energy (CAPE) time scale (Appendix~\ref{sec:rc_setup}), which is a key parameter controlling the amount of parameterized convective mass flux relative to the resolved vertical motion and associated rainfall in the model \citep{Tomassini23_confronting}.
In summary, stretching the mesh while keeping the convection parameterization the same results in relatively more grid-scale precipitation; while reducing or disabling this parameterization offloads the precipitation to the grid scale.

\section{Discussion}
\label{sec:discussion}
Our results show that LFRic-Atmosphere is capable of simulating small-scale atmospheric processes using a stretched mesh while also reproducing the key characteristics of the global climate of a tidally locked terrestrial exoplanet.
Using a locally refined mesh allows us to capture small-scale cloud patterns in the substellar region.
In agreement with previous studies based on localized or global high-resolution grids \citep[e.g.][]{Harris16_high-resolution, Kajikawa16_resolution, Uchida16_error, Rios-Berrios22_differences}, resolving deep convection leads to more intense and localized precipitation.
Thus, at a reduced computational cost (Table~\ref{tab:tech}), LFRic-Atmosphere allows us to examine a regional phenomenon such as convection in great detail and at the same time study the interaction between local and global atmospheric circulation.

Tidally locked exoplanets offer a particularly suitable use case but a locally refined mesh has also shown potential for Mars climate modeling, e.g. at the scale of individual craters \citep{Lian23_unstructured}.
Stretched-mesh GCMs may also be useful in studying Titan weather, in particular the interaction between localized convective storms and large-scale Rossby waves \citep{Battalio21_global}.
For Venus and slowly-rotating Venus-like exoplanets, stretched-mesh GCMs may shed light on the convectively-generated gravity waves and their interaction with the superrotating flow \citep{Lefevre18_three-dimensional}.
For non tidally locked exoplanets, the mesh can be stretched in a different way, e.g. focusing the high resolution on the equatorial band, allowing for a better simulation of deep convection in the tropics \citep{Rios-Berrios22_differences}.

Our study is the first exoplanet application of LFRic-Atmosphere in a global configuration with explicit convection, allowed for by the stretched mesh.
Our key finding is that the simulation with explicit convection (\expSE) predicts a climate that is substantially drier (as seen in both specific and relative humidity) and colder (by \SI{\approx 5}{\K}), with about \SI{10}{\percent} more cloud condensate compared to the simulations with fully parameterized convection.
The main reason for the colder climate is the higher cloud albedo, i.e. stronger reflection of the shortwave radiation by clouds (by \SI{\approx 10}{\watt\per\m\squared}).
Additionally, there is less water vapor and thus a weaker greenhouse effect.
The increase in cloud condensate, especially in the low-level cloud liquid, can be partly explained by the fact that the grid resolution on the day side's periphery may be too coarse to fully resolve the shallow convection.
However, the experiment with reduced parameterization (\expSR) produces a similar cloud structure (between that in the \expSE run and those with fully parameterized convection), supporting this trend.
Note that the \expSR simulation serves as an intermediate solution between the parameterized and explicit convection because it is designed to activate convection depending not only on the grid cell size but the size of convection relative to it.

We believe that overall our experiment with explicit convection simulates a climate close to ``reality'' because it truly represents the underlying physics of the atmosphere, without relying on parameterizations tuned for Earth.
The main caveat is the grid coarsening on the day-side periphery where convection becomes under-resolved.
A possible solution could be to use a base mesh with more grid points and a smaller stretching factor.
In the absence of observational constraints, these simulations may serve as benchmarks for coarse-resolution GCMs with parameterized convection.
At the same time, as noted in \citet{Tomassini23_confronting}, cloud parameterizations will likely require tuning for convection-permitting models.
This may reduce the amount of shortwave radiation reflected by the clouds, resulting in a warmer climate.
Future work should explore the representation of cloud microphysics and their impact on the climate in stretched-mesh simulations.

Another caveat of the present study is that the convection scheme in our model is not tuned for a stretched mesh configuration.
Indeed, in most km-scale models convection is only partially resolved \citep{Tomassini23_confronting}.
This introduces the challenge of representing convection in a balanced way: partially explicitly and partially as a parameterized sub-grid process.
Scale-aware convection schemes are a promising solution \citep{Kendon21_challenges}, which may also re-ignite interest in using stretched-mesh GCMs \citep{Uchida16_error}.
One example of a scale-aware scheme is CoMorph, a new convection parametrization under development at the Met Office \citep{Daleu23_evaluating, Lavender24_use}.
It is designed for use within both the Unified Model and LFRic-Atmosphere and has already shown to be close to cloud-resolving models in representing many aspects of convection \citep{Lavender24_use}.
The benefits of applying CoMorph in a stretched-mesh configuration will be explored for a tidally locked exoplanet in a future study.

Despite the differences discussed above, the key climate diagnostics and spatial patterns simulated in the stretched and non-stretched runs are relatively similar and the fidelity of the large-scale atmospheric circulation is preserved.
We have thus shown that mesh stretching allows for a gradual transition into high resolution and does not produce numerical artifacts, confirming previous work on Earth climate \citep{Fox-Rabinovitz00_uniform, Harris16_high-resolution, Uchida16_error} and opening new avenues for studying 3D mixing in planetary atmospheres.

We thank Lorenzo Tomassini for his help with the reduced convection setup.
We thank Thomas Melvin and Nigel Wood for helpful feedback on this work.
We thank Thomas Melvin for his help with the cubed sphere mesh visualization.
Material produced using Met Office Software.
We acknowledge use of the Monsoon2 system, a collaborative facility supplied under the Joint Weather and Climate Research Programme, a strategic partnership between the Met Office and the Natural Environment Research Council.
Additionally, some of this work was performed using the DiRAC Data Intensive service at Leicester, operated by the University of Leicester IT Services, which forms part of the STFC DiRAC HPC Facility (\url{www.dirac.ac.uk}).
The equipment was funded by BEIS capital funding via STFC capital grants \texttt{ST/K000373/1} and \texttt{ST/R002363/1} and STFC DiRAC Operations grant \texttt{ST/R001014/1}.
DiRAC is part of the National e-Infrastructure.
This work was supported by a UKRI Future Leaders Fellowship \texttt{MR/T040866/1}.
This work also was partly funded by the Leverhulme Trust through a research project grant \texttt{RPG-2020-82}.

\software{
The LFRic-Atmosphere source code and configuration files are freely available from the Met Office Science Repository Service (\url{https://code.metoffice.gov.uk}) upon registration and completion of a software license.
The UM and JULES code used in the publication has been committed to the UM and JULES code trunks, having passed both science and code reviews according to the UM and JULES working practices; in the UM/JULES versions stated in the paper (\texttt{vn13.3}).
Scripts to post-process and visualize model output are available at \url{https://github.com/dennissergeev/stretched_mesh_code} and depend on the following open-source Python libraries: \texttt{aeolus} \citep{Sergeev24_aeolus}, \texttt{geovista} \citep{Little23_geovista}, \texttt{iris-esmf-regrid} \citep{Worsley23_irisesmfregrid}, \texttt{iris} \citep{iris}, \texttt{matplotlib} \citep{Hunter07_matplotlib}, and \texttt{numpy} \citep{Harris20_array}.
          }

\appendix
\section{Mesh Stretching} 
\label{sec:mesh_stretch}

The stretching of the mesh is achieved using the \citet{Schmidt77_variable} transformation \citep[see also][]{Harris16_high-resolution, Bindle21_grid-stretching}.
It preserves the mesh topology, so the number of grid points and their connectivity are the same as in the non-stretched mesh (Fig.~\ref{fig:mesh_geom}d).
The stretching is smooth and does not have abrupt transitions in resolution such as those associated with grid nesting \citep{Sergeev20_atmospheric}.
The resulting grid has a refined resolution at the target location compensated by a coarser resolution on the opposite side of the sphere.
Finally, the refined domain (the `target face') diminishes in size proportional to the stretching factor.

The mesh stretching procedure involves two steps.
First, the grid points of the original cubed sphere are attracted to the South Pole by the \citet{Schmidt77_variable} transform
$$\phi'(\phi) = \arcsin\left(\frac{D+\sin\phi}{1+D\sin\phi}\right)~\text{with}~D=\frac{1-S^2}{1+S^2},$$
where $S$ is the stretch factor ($S>1$ causes stretching), $\phi$ is the original latitude, and $\phi'$ is the transformed latitude.
The second step is the rotation of the stretched mesh so that the refined region is centered at the desired location.
Thus, four parameters uniquely define a stretched mesh: the size of the cubed sphere C$n$, the stretch factor $S$, the target latitude $\phi_T$ and longitude $\lambda_T$.
The larger the $S$ factor, the finer and more localized the refinement is.
The grid spacing at the target location is approximately $S$ times finer compared to the non-stretched mesh, while at the antipode location it is approximately $S$ times coarser.
Stretch factors used in existing literature are typically between 2.5 and 10 \citep[e.g.][]{Fox-Rabinovitz06_variable, Harris16_high-resolution, Uchida16_error, Bindle21_grid-stretching}.

As shown in Fig.~\ref{fig:mesh_geom}, we use a stretched mesh with $s=10$ which gives a resolution $\approx 10$$\times$ finer at the target location.
The target location is the substellar point $\phi_T=\ang{0}$,  $\lambda_T=\ang{0}$.
While the factor of 10 is high, it has been used for Earth applications, e.g. to better resolve nitrogen dioxide emissions in California \citep{Bindle21_grid-stretching}.
By using a highly stretched mesh, we also test the boundaries of the computational stability of LFRic-Atmosphere.
In future studies, we will probe other stretch factors, depending on the planet and problem in question.

The resulting stretched mesh has a cell width size of \SI{\approx 4.7}{\km} in the substellar region (Fig.~\ref{fig:mesh_geom}a--c), reaching the convection-permitting resolution.
To achieve a similar resolution with a quasi-uniform mesh, one would have to use 100 times as many grid points, which is extremely computationally expensive for climate simulations.
Similar 4-km grid spacing has been used for simulating convection explicitly in regional climate predictions for Earth \citep[e.g.][]{Stratton18_pan-african} and, more recently, for exoplanets with an Earth-like atmosphere, both on a global \citep{Yang23_cloud} and regional \citep{Sergeev20_atmospheric} scale.
By the nature of grid stretching, the night side of the planet has a relatively coarse mesh spacing of up to \SI{\approx 470}{\km}.
Nevertheless, even this resolution is close to those applied \textit{globally} in previous studies for exoplanetary climates \citep[e.g.][]{Yang13_stabilizing, Turbet16_habitability, Paradise22_exoplasim}.

As Table~\ref{tab:tech} shows, the cost of performing stretched-mesh simulations is only about 2$\times$ higher than that of the uniform-mesh simulation (mostly due to a smaller time step in the stretched-mesh setup), while allowing us to reach a convection-permitting resolution for the substellar point.
Using a uniform mesh with the same high resolution for the whole planet would require a relative computational cost of $\sim$1000.

\begin{deluxetable*}{lllll}
\tablecaption{Technical details of the experiments.\label{tab:tech}}
\tablewidth{0pt}
\tablehead{
& \colhead{\expSE} & \colhead{\expSR} &\colhead{\expSP} & \colhead{\expUP}
}
\startdata
Mesh & C192 & & \\
Number of columns & 221184 & & & \\
Stretch factor $S$ & 10 & 10 & 10 & 1 (no stretching) \\
Number of vertical levels & 63 & & & \\
Model top (km) & 41.02239 & & & \\ 
Dynamics time step (seconds) & 120 & 120 & 120 & 300 \\
Radiative transfer time step (seconds) & 1200 & & & \\
& & & & \\
System configuration & \multicolumn{4}{l}{Cray XC-40 Broadwell Nodes (36 CPUs per node)} \\
Number of cores & 864 & & & \\
Relative wall time per model time & 1.92 & 2.09 & 2.18 & 1.00 \\
Relative throughput (model time per wall time) & 0.52 & 0.48 & 0.46 & 1.00\\
\enddata
\end{deluxetable*}

\section{Convection and turbulence parameters in the \expSR experiment}
\label{sec:rc_setup}

In the \expSR experiment, we follow a km-scale configuration used for the Unified Model in \citet{Tomassini23_confronting}, namely their \emph{MidLevShConv15RAturb} configuration.
This configuration combines a ``reduced'' convection parameterization with a modified representation boundary layer scheme.
As a result, convection is allowed to be partly explicit, accounting for the fact that in the focal region of our stretched mesh the grid step reaches \SI{4.7}{\km}.

Briefly, the \expSR parameterization settings are different to those used in \expSP and \expUP in the following. 
Scaling of the shallow convective mass flux is set to 0.015 (default: 0.03) --- the value found to produce more realistic low-level clouds in global km-scale numerical weather prediction runs.
The formulation of the mid-level convection scheme is also modified.
The key modification that reduces the sub-grid convective mass flux is a longer timescale for the convective available potential energy (CAPE) closure: \SI{2700}{\s} instead of the default \SI{1800}{\s}.
This effectively decreases the strength of the convection parameterization and allows convection to be partly explicit, i.e. represented by grid-scale vertical flow.
The configuration also includes the turbulence blending scheme \citep{Boutle14_seamless}, which allows for a transition between the 1D turbulence parameterization (used in coarse-grid GCMs) with a 3D Smagorinsky-Lilly turbulence scheme (used in cloud-resolving models).
As mentioned in Sec.~\ref{sec:results}, this parameterization leads to a drier stratosphere compared to the other three simulations.

\section{Circulation regime bistability}
\label{sec:bistability}
\begin{figure*}
    \centering
    \includegraphics[width=1\linewidth]{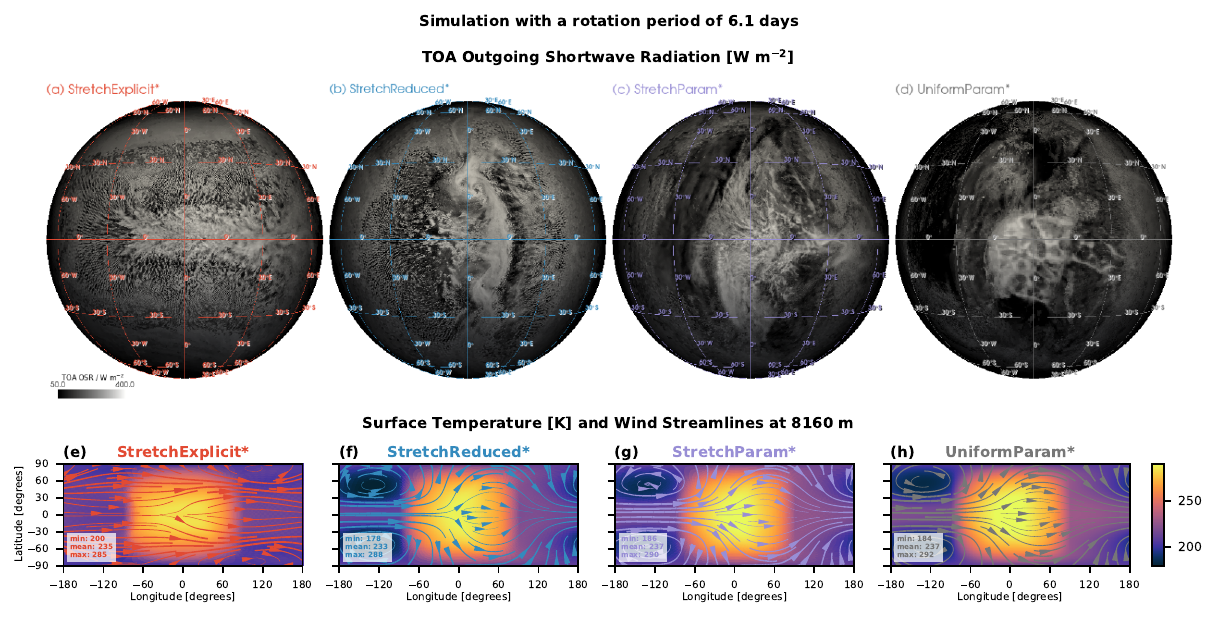}
    \caption{
    Circulation regime bistability in the simulations with the rotation period set to 6.1 days.
    The circulation regime switches to another state when the convection scheme is switched off --- compare the two leftmost columns (the \expSE and \expSR cases).
    Top panels show instantaneous top-of-atmosphere outgoing shortwave radiation.
    Bottom panels show the time mean surface temperature in \si{\K} overlaid by streamlines of the horizontal wind at \SI{\approx 8}{\km} above the surface.
    }
    \label{fig:thai_hab1_orig}
\end{figure*}

In additional experiments, we change the rotation rate of TRAPPIST-1e to show the climate bistability.
We set it to its observed orbital period of 6.1 days, i.e. the value prescribed by the THAI protocol \citep{Fauchez20_thai_protocol} and used in the majority of modeling studies for this planet \citep[e.g.][]{Wolf17_assessing, Turbet18_modeling, Fauchez19_impact, Eager-Nash20_implications, May21_water, Turbet22_thai, Yang23_cloud}.
Broadly, our conclusions hold for these simulations.
The key difference is that in this case the GCM is prone to a climate bistability: the atmospheric circulation settles on a double-jet regime in the \expSE experiment, and on a single-jet regime in the other three experiments \citep[for definitions, see][]{Sergeev22_bistability}.
This corresponds to a distinct difference in climate characteristics, including the global cloud cover (Fig.~\ref{fig:thai_hab1_orig}).
In the \expSE experiment, the cloud cover associated with the deep convection is narrower and more zonally elongated than that in the other three experiments.

The fact that the circulation regime changes not when the mesh is stretched (comparing \expUP to \expSP) but when the convection scheme is disabled (comparing \expSP to \expSE), confirms our key finding that for the global climate mesh stretching is less important than changes in model parameterizations (see Sec.~\ref{sec:results}).
Furthermore, our \expSE experiment is qualitatively similar to the convection-permitting simulation of TRAPPIST-1e in \citet{Yang23_cloud}, in terms of the spatial distribution of temperature, wind, cloud cover, and precipitation.
This implies that at least some of the differences between the coarse-grid GCMs and the convection-permitting model reported by \citeauthor{Yang23_cloud} are due to the circulation regime change and not due to resolving convection per se.
And as \citet{Sergeev22_bistability} demonstrated, the circulation regime change can happen even in a coarse-grid GCM if the convection parameterization is modified or switched off.

Despite the circulation regime change in these simulations, our model predicts a climate within the spread of THAI GCM results \citep{Sergeev22_thai}.
Taking the experiment with the uniform mesh (\expUP) as an example, the global mean surface temperature is \SI{237}{\K} (Fig.~\ref{fig:thai_hab1_orig}h), which is lower than that predicted by ExoCAM and LMD-G but higher than that predicted by the UM for the Hab~1 case.
Likewise, the maximum surface temperature (\SI{292}{\K}) falls within the inter-model spread.
The lowest surface temperature is \SI{184}{\K}, simulated in the regions of night-side stationary gyres.
It is on the colder end of the spectrum of the THAI GCMs but still about \SI{10}{\K} higher than that in the UM simulation.
Comparing the \expSE case with the ROCKE-3D simulation in \citet{Sergeev22_thai}, we see the same trend in the surface temperature in both models: the minimum temperatures rise to about \SI{200}{\K} due to the shift of the stationary waves, while the maximum temperatures fall to \SIrange{285}{290}{\K}.
Thus, in the THAI Hab~1 case, LFRic-Atmosphere tends to produce a colder climate compared to other GCMs, though not as cold as that in the UM.

\bibliography{references}{}

\begin{thebibliography}{}
\expandafter\ifx\csname natexlab\endcsname\relax\def\natexlab#1{#1}\fi
\providecommand{\url}[1]{\href{#1}{#1}}
\providecommand{\dodoi}[1]{doi:~\href{http://doi.org/#1}{\nolinkurl{#1}}}
\providecommand{\doeprint}[1]{\href{http://ascl.net/#1}{\nolinkurl{http://ascl.net/#1}}}
\providecommand{\doarXiv}[1]{\href{https://arxiv.org/abs/#1}{\nolinkurl{https://arxiv.org/abs/#1}}}

\bibitem[{Adams {et~al.}(2019)Adams, Ford, Hambley, Hobson, Kavčič, Maynard,
  Melvin, Müller, Mullerworth, Porter, Rezny, Shipway, \&
  Wong}]{Adams19_lfric}
Adams, S., Ford, R., Hambley, M., {et~al.} 2019, Journal of Parallel and
  Distributed Computing, 132, 383, \dodoi{10.1016/j.jpdc.2019.02.007}

\bibitem[{Arakawa(2004)}]{Arakawa04_cumulus}
Arakawa, A. 2004, Journal of Climate, 17, 2493,
  \dodoi{10.1175/1520-0442(2004)017<2493:RATCPP>2.0.CO;2}

\bibitem[{Battalio \& Lora(2021)}]{Battalio21_global}
Battalio, J.~M., \& Lora, J.~M. 2021, Geophysical Research Letters, 48,
  e2021GL094244, \dodoi{10.1029/2021GL094244}

\bibitem[{Bendall {et~al.}(2020)Bendall, Gibson, Shipton, Cotter, \&
  Shipway}]{Bendall20_compatible}
Bendall, T.~M., Gibson, T.~H., Shipton, J., Cotter, C.~J., \& Shipway, B. 2020,
  Quarterly Journal of the Royal Meteorological Society, 146, 3187,
  \dodoi{10.1002/qj.3841}

\bibitem[{Bendall {et~al.}(2022)Bendall, Wood, Thuburn, \&
  Cotter}]{Bendall22_solution}
Bendall, T.~M., Wood, N., Thuburn, J., \& Cotter, C.~J. 2022, Quarterly Journal
  of the Royal Meteorological Society, \dodoi{10.1002/qj.4406}

\bibitem[{Bindle {et~al.}(2021)Bindle, Martin, Cooper, Lundgren, Eastham, Auer,
  Clune, Weng, Lin, Murray, Meng, Keller, Putman, Pawson, \&
  Jacob}]{Bindle21_grid-stretching}
Bindle, L., Martin, R.~V., Cooper, M.~J., {et~al.} 2021, Geoscientific Model
  Development, 14, 5977, \dodoi{10.5194/gmd-14-5977-2021}

\bibitem[{Boutle {et~al.}(2014)Boutle, Eyre, \& Lock}]{Boutle14_seamless}
Boutle, I.~A., Eyre, J. E.~J., \& Lock, A.~P. 2014, Monthly Weather Review,
  142, 1655, \dodoi{10.1175/MWR-D-13-00229.1}

\bibitem[{Brown {et~al.}(2023)Brown, Bendall, Boutle, Melvin, \&
  Shipway}]{Brown23_physicsdynamicschemistry}
Brown, A., Bendall, T.~M., Boutle, I., Melvin, T., \& Shipway, B. 2023,
  Physics-{Dynamics}-{Chemistry} {Coupling} {Across} {Different} {Meshes} in
  {LFRic}-{Atmosphere}: {Formulation} and {Idealised} {Tests},  arXiv,
  \dodoi{10.48550/arXiv.2310.01255}

\bibitem[{Daleu {et~al.}(2023)Daleu, Plant, Stirling, \&
  Whitall}]{Daleu23_evaluating}
Daleu, C.~L., Plant, R.~S., Stirling, A.~J., \& Whitall, M. 2023, Quarterly
  Journal of the Royal Meteorological Society, n/a, \dodoi{10.1002/qj.4547}

\bibitem[{Eager-Nash {et~al.}(2020)Eager-Nash, Reichelt, Mayne, Lambert,
  Sergeev, Ridgway, Manners, Boutle, Lenton, \&
  Kohary}]{Eager-Nash20_implications}
Eager-Nash, J.~K., Reichelt, D.~J., Mayne, N.~J., {et~al.} 2020, Astronomy \&
  Astrophysics, 639, A99, \dodoi{10.1051/0004-6361/202038089}

\bibitem[{Edson {et~al.}(2011)Edson, Lee, Bannon, Kasting, \&
  Pollard}]{Edson11_atmospheric}
Edson, A., Lee, S., Bannon, P., Kasting, J.~F., \& Pollard, D. 2011, Icarus,
  212, 1, \dodoi{10.1016/j.icarus.2010.11.023}

\bibitem[{Edwards \& Slingo(1996)}]{Edwards96_studies}
Edwards, J.~M., \& Slingo, A. 1996, Quarterly Journal of the Royal
  Meteorological Society, 122, 689, \dodoi{10.1002/qj.49712253107}

\bibitem[{Fauchez {et~al.}(2019)Fauchez, Turbet, Villanueva, Wolf, Arney,
  Kopparapu, Lincowski, Mandell, Wit, Pidhorodetska, Domagal-Goldman, \&
  Stevenson}]{Fauchez19_impact}
Fauchez, T.~J., Turbet, M., Villanueva, G.~L., {et~al.} 2019, The Astrophysical
  Journal, 887, 194, \dodoi{10.3847/1538-4357/ab5862}

\bibitem[{Fauchez {et~al.}(2020)Fauchez, Turbet, Wolf, Boutle, Way, Del~Genio,
  Mayne, Tsigaridis, Kopparapu, Yang, Forget, Mandell, \&
  Domagal~Goldman}]{Fauchez20_thai_protocol}
Fauchez, T.~J., Turbet, M., Wolf, E.~T., {et~al.} 2020, Geosci. Model Dev, 13,
  707, \dodoi{10.5194/gmd-13-707-2020}

\bibitem[{Fauchez {et~al.}(2022)Fauchez, Villanueva, Sergeev, Turbet, Boutle,
  Tsigaridis, Way, Wolf, Domagal-Goldman, Forget, Haqq-Misra, Kopparapu,
  Manners, \& Mayne}]{Fauchez22_thai}
Fauchez, T.~J., Villanueva, G.~L., Sergeev, D.~E., {et~al.} 2022, The Planetary
  Science Journal, 3, 213, \dodoi{10.3847/PSJ/ac6cf1}

\bibitem[{Fox-Rabinovitz {et~al.}(2008)Fox-Rabinovitz, Cote, Dugas, Deque,
  McGregor, \& Belochitski}]{Fox-Rabinovitz08_stretchedgrid}
Fox-Rabinovitz, M., Cote, J., Dugas, B., {et~al.} 2008, Meteorology and
  Atmospheric Physics, 100, 159, \dodoi{10.1007/s00703-008-0301-z}

\bibitem[{Fox-Rabinovitz {et~al.}(2006)Fox-Rabinovitz, Côté, Dugas, Déqué,
  \& McGregor}]{Fox-Rabinovitz06_variable}
Fox-Rabinovitz, M., Côté, J., Dugas, B., Déqué, M., \& McGregor, J.~L.
  2006, Journal of Geophysical Research: Atmospheres, 111,
  \dodoi{10.1029/2005JD006520}

\bibitem[{Fox-Rabinovitz(2000)}]{Fox-Rabinovitz00_simulation}
Fox-Rabinovitz, M.~S. 2000, Journal of Geophysical Research: Atmospheres, 105,
  29635, \dodoi{10.1029/2000JD900650}

\bibitem[{Fox-Rabinovitz {et~al.}(2000)Fox-Rabinovitz, Stenchikov, Suarez,
  Takacs, \& Govindaraju}]{Fox-Rabinovitz00_uniform}
Fox-Rabinovitz, M.~S., Stenchikov, G.~L., Suarez, M.~J., Takacs, L.~L., \&
  Govindaraju, R.~C. 2000, Monthly Weather Review, 128, 1883,
  \dodoi{10.1175/1520-0493(2000)128<1883:AUAVRS>2.0.CO;2}

\bibitem[{Gryschka \& Raasch(2005)}]{Gryschka05_roll}
Gryschka, M., \& Raasch, S. 2005, Geophysical Research Letters, 32, 1,
  \dodoi{10.1029/2005GL022872}

\bibitem[{Hammond {et~al.}(2020)Hammond, Tsai, \&
  Pierrehumbert}]{Hammond20_equatorial}
Hammond, M., Tsai, S.-M., \& Pierrehumbert, R.~T. 2020, The Astrophysical
  Journal, 901, 78, \dodoi{10.3847/1538-4357/abb08b}

\bibitem[{Haqq-Misra {et~al.}(2018)Haqq-Misra, Wolf, Joshi, Zhang, \&
  Kopparapu}]{Haqq-Misra18_demarcating}
Haqq-Misra, J., Wolf, E.~T., Joshi, M., Zhang, X., \& Kopparapu, R.~K. 2018,
  The Astrophysical Journal, 852, 67, \dodoi{10.3847/1538-4357/aa9f1f}

\bibitem[{Harris {et~al.}(2020)Harris, Millman, van~der Walt, Gommers,
  Virtanen, Cournapeau, Wieser, Taylor, Berg, Smith, Kern, Picus, Hoyer, van
  Kerkwijk, Brett, Haldane, del Río, Wiebe, Peterson, Gérard-Marchant,
  Sheppard, Reddy, Weckesser, Abbasi, Gohlke, \& Oliphant}]{Harris20_array}
Harris, C.~R., Millman, K.~J., van~der Walt, S.~J., {et~al.} 2020, Nature, 585,
  357, \dodoi{10.1038/s41586-020-2649-2}

\bibitem[{Harris {et~al.}(2016)Harris, Lin, \& Tu}]{Harris16_high-resolution}
Harris, L.~M., Lin, S.-J., \& Tu, C. 2016, Journal of Climate, 29, 4293,
  \dodoi{10.1175/JCLI-D-15-0389.1}

\bibitem[{Hunter(2007)}]{Hunter07_matplotlib}
Hunter, J.~D. 2007, Computing in Science \& Engineering, 9, 90,
  \dodoi{10.1109/MCSE.2007.55}

\bibitem[{Jenney {et~al.}(2023)Jenney, Ferretti, \&
  Pritchard}]{Jenney23_vertical}
Jenney, A.~M., Ferretti, S.~L., \& Pritchard, M.~S. 2023, Journal of Advances
  in Modeling Earth Systems, 15, e2022MS003444, \dodoi{10.1029/2022MS003444}

\bibitem[{Kajikawa {et~al.}(2016)Kajikawa, Miyamoto, Yoshida, Yamaura, Yashiro,
  \& Tomita}]{Kajikawa16_resolution}
Kajikawa, Y., Miyamoto, Y., Yoshida, R., {et~al.} 2016, Progress in Earth and
  Planetary Science, 3, 16, \dodoi{10.1186/s40645-016-0094-5}

\bibitem[{Kendon {et~al.}(2021)Kendon, Prein, Senior, \&
  Stirling}]{Kendon21_challenges}
Kendon, E.~J., Prein, A.~F., Senior, C.~A., \& Stirling, A. 2021, Philosophical
  Transactions of the Royal Society A: Mathematical, Physical and Engineering
  Sciences, 379, 20190547, \dodoi{10.1098/rsta.2019.0547}

\bibitem[{Kent {et~al.}(2023)Kent, Melvin, \& Wimmer}]{Kent23_mixed}
Kent, J., Melvin, T., \& Wimmer, G.~A. 2023, Geoscientific Model Development,
  16, 1265, \dodoi{10.5194/gmd-16-1265-2023}

\bibitem[{Komacek \& Abbot(2019)}]{Komacek19_atmospheric}
Komacek, T.~D., \& Abbot, D.~S. 2019, The Astrophysical Journal, 871, 245,
  \dodoi{10.3847/1538-4357/aafb33}

\bibitem[{Lavender {et~al.}(2024)Lavender, Stirling, Whitall, Stratton, Daleu,
  Plant, Lock, \& Gu}]{Lavender24_use}
Lavender, S.~L., Stirling, A.~J., Whitall, M., {et~al.} 2024, Quarterly Journal
  of the Royal Meteorological Society, n/a, \dodoi{10.1002/qj.4660}

\bibitem[{Lefèvre {et~al.}(2018)Lefèvre, Lebonnois, \&
  Spiga}]{Lefevre18_three-dimensional}
Lefèvre, M., Lebonnois, S., \& Spiga, A. 2018, Journal of Geophysical
  Research: Planets, 123, 2773, \dodoi{10.1029/2018JE005679}

\bibitem[{Lefèvre {et~al.}(2021)Lefèvre, Turbet, \&
  Pierrehumbert}]{Lefevre21_3d}
Lefèvre, M., Turbet, M., \& Pierrehumbert, R. 2021, The Astrophysical Journal,
  913, 101, \dodoi{10.3847/1538-4357/abf2c1}

\bibitem[{Lian \& Richardson(2023)}]{Lian23_unstructured}
Lian, Y., \& Richardson, M.~I. 2023, Planetary and Space Science, 229, 105663,
  \dodoi{10.1016/j.pss.2023.105663}

\bibitem[{Little(2023)}]{Little23_geovista}
Little, B. 2023, geovista, \dodoi{10.5281/zenodo.7608302}

\bibitem[{May {et~al.}(2021)May, Taylor, Komacek, Line, \&
  Parmentier}]{May21_water}
May, E.~M., Taylor, J., Komacek, T.~D., Line, M.~R., \& Parmentier, V. 2021,
  The Astrophysical Journal Letters, 911, L30, \dodoi{10.3847/2041-8213/abeeff}

\bibitem[{Melvin {et~al.}(2019)Melvin, Benacchio, Shipway, Wood, Thuburn, \&
  Cotter}]{Melvin19_mixed}
Melvin, T., Benacchio, T., Shipway, B., {et~al.} 2019, Quarterly Journal of the
  Royal Meteorological Society, 145, 2835, \dodoi{10.1002/QJ.3501}

\bibitem[{Melvin {et~al.}(2024)Melvin, Shipway, Wood, Benacchio, Bendall,
  Boutle, Brown, Johnson, Kent, Pring, Smith, Zerroukat, Cotter, \&
  Thuburn}]{Melvin24_mixed}
Melvin, T., Shipway, B., Wood, N., {et~al.} 2024, A mixed finite-element,
  finite-volume, semi-implicit discretisation for atmospheric dynamics:
  {Spherical} geometry,  arXiv, \dodoi{10.48550/arXiv.2402.13738}

\bibitem[{{Met Office}(2023)}]{iris}
{Met Office}. 2023, Iris: {A} powerful, format-agnostic, and community-driven
  {Python} package for analysing and visualising {Earth} science data,
  \dodoi{10.5281/zenodo.7948293}

\bibitem[{Paradise {et~al.}(2022)Paradise, Macdonald, Menou, Lee, \&
  Fan}]{Paradise22_exoplasim}
Paradise, A., Macdonald, E., Menou, K., Lee, C., \& Fan, B.~L. 2022, Monthly
  Notices of the Royal Astronomical Society, 511, 3272,
  \dodoi{10.1093/mnras/stac172}

\bibitem[{Rios‐Berrios {et~al.}(2022)Rios‐Berrios, Bryan, Medeiros, Judt,
  \& Wang}]{Rios-Berrios22_differences}
Rios‐Berrios, R., Bryan, G.~H., Medeiros, B., Judt, F., \& Wang, W. 2022,
  Journal of Advances in Modeling Earth Systems, 14, e2021MS002902,
  \dodoi{10.1029/2021MS002902}

\bibitem[{Schmidt(1977)}]{Schmidt77_variable}
Schmidt, F. 1977, Contributions to Atmospheric Physics, 50, 211,
  \dodoi{10.1186/s40645-014-0018-1.Schmidt}

\bibitem[{Sergeev {et~al.}(2020)Sergeev, Lambert, Mayne, Boutle, Manners, \&
  Kohary}]{Sergeev20_atmospheric}
Sergeev, D.~E., Lambert, F.~H., Mayne, N.~J., {et~al.} 2020, The Astrophysical
  Journal, 894, 84, \dodoi{10.3847/1538-4357/ab8882}

\bibitem[{Sergeev {et~al.}(2022{\natexlab{a}})Sergeev, Lewis, Lambert, Mayne,
  Boutle, Manners, \& Kohary}]{Sergeev22_bistability}
Sergeev, D.~E., Lewis, N.~T., Lambert, F.~H., {et~al.} 2022{\natexlab{a}}, The
  Planetary Science Journal, 3, 214, \dodoi{10.3847/PSJ/ac83be}

\bibitem[{Sergeev \& Zamyatina(2024)}]{Sergeev24_aeolus}
Sergeev, D.~E., \& Zamyatina, M. 2024, Aeolus - a {Python} library for the
  analysis and visualisation of climate model output.,
  \dodoi{10.5281/zenodo.5145603}

\bibitem[{Sergeev {et~al.}(2022{\natexlab{b}})Sergeev, Fauchez, Turbet, Boutle,
  Tsigaridis, Way, Wolf, Domagal-Goldman, Forget, Haqq-Misra, Kopparapu,
  Lambert, Manners, \& Mayne}]{Sergeev22_thai}
Sergeev, D.~E., Fauchez, T.~J., Turbet, M., {et~al.} 2022{\natexlab{b}}, The
  Planetary Science Journal, 3, 212, \dodoi{10.3847/PSJ/ac6cf2}

\bibitem[{Sergeev {et~al.}(2023)Sergeev, Mayne, Bendall, Boutle, Brown,
  Kavčič, Kent, Kohary, Manners, Melvin, Olivier, Ragta, Shipway, Wakelin,
  Wood, \& Zerroukat}]{Sergeev23_simulations}
Sergeev, D.~E., Mayne, N.~J., Bendall, T., {et~al.} 2023, Geoscientific Model
  Development, 16, 5601, \dodoi{10.5194/gmd-16-5601-2023}

\bibitem[{Stratton {et~al.}(2018)Stratton, Senior, Vosper, Folwell, Boutle,
  Earnshaw, Kendon, Lock, Malcolm, Manners, Morcrette, Short, Stirling, Taylor,
  Tucker, Webster, \& Wilkinson}]{Stratton18_pan-african}
Stratton, R.~A., Senior, C.~A., Vosper, S.~B., {et~al.} 2018, Journal of
  Climate, 31, 3485, \dodoi{10.1175/JCLI-D-17-0503.1}

\bibitem[{Tomassini {et~al.}(2023)Tomassini, Willett, Sellar, Lock, Walters,
  Whitall, Sanchez, Heming, Earnshaw, Rodriguez, Ackerley, Xavier, Franklin, \&
  Senior}]{Tomassini23_confronting}
Tomassini, L., Willett, M., Sellar, A., {et~al.} 2023, Journal of Advances in
  Modeling Earth Systems, 15, e2022MS003418, \dodoi{10.1029/2022MS003418}

\bibitem[{Turbet {et~al.}(2016)Turbet, Leconte, Selsis, Bolmont, Forget, Ribas,
  Raymond, \& Anglada-Escudé}]{Turbet16_habitability}
Turbet, M., Leconte, J., Selsis, F., {et~al.} 2016, Astronomy \& Astrophysics,
  596, A112, \dodoi{10.1051/0004-6361/201629577}

\bibitem[{Turbet {et~al.}(2018)Turbet, Bolmont, Leconte, Forget, Selsis, Tobie,
  Caldas, Naar, \& Gillon}]{Turbet18_modeling}
Turbet, M., Bolmont, E., Leconte, J., {et~al.} 2018, Astronomy \& Astrophysics,
  612, A86, \dodoi{10.1051/0004-6361/201731620}

\bibitem[{Turbet {et~al.}(2022)Turbet, Fauchez, Sergeev, Boutle, Tsigaridis,
  Way, Wolf, Domagal-Goldman, Forget, Haqq-Misra, Kopparapu, Lambert, Manners,
  Mayne, \& Sohl}]{Turbet22_thai}
Turbet, M., Fauchez, T.~J., Sergeev, D.~E., {et~al.} 2022, The Planetary
  Science Journal, 3, 211, \dodoi{10.3847/PSJ/ac6cf0}

\bibitem[{Uchida {et~al.}(2016)Uchida, Mori, Nakamura, Satoh, Suzuki, \&
  Nakajima}]{Uchida16_error}
Uchida, J., Mori, M., Nakamura, H., {et~al.} 2016, Monthly Weather Review, 144,
  1423, \dodoi{10.1175/MWR-D-15-0271.1}

\bibitem[{Walters {et~al.}(2019)Walters, Baran, Boutle, Brooks, Earnshaw,
  Edwards, Furtado, Hill, Lock, Manners, Morcrette, Mulcahy, Sanchez, Smith,
  Stratton, Tennant, Tomassini, Van~Weverberg, Vosper, Willett, Browse,
  Bushell, Carslaw, Dalvi, Essery, Gedney, Hardiman, Johnson, Johnson, Jones,
  Jones, Mann, Milton, Rumbold, Sellar, Ujiie, Whitall, Williams, \&
  Zerroukat}]{Walters19_ga7}
Walters, D., Baran, A.~J., Boutle, I., {et~al.} 2019, Geoscientific Model
  Development, 12, 1909, \dodoi{10.5194/gmd-12-1909-2019}

\bibitem[{Wei {et~al.}(2020)Wei, Zhang, \& Yang}]{Wei20_small}
Wei, M., Zhang, Y., \& Yang, J. 2020, The Astrophysical Journal, 898, 156,
  \dodoi{10.3847/1538-4357/ab9b83}

\bibitem[{Wolf(2017)}]{Wolf17_assessing}
Wolf, E.~T. 2017, The Astrophysical Journal, 839, L1,
  \dodoi{10.3847/2041-8213/aa693a}

\bibitem[{Wolf {et~al.}(2022)Wolf, Kopparapu, Haqq-Misra, \&
  Fauchez}]{Wolf22_exocam}
Wolf, E.~T., Kopparapu, R., Haqq-Misra, J., \& Fauchez, T.~J. 2022, The
  Planetary Science Journal, 3, 7, \dodoi{10.3847/PSJ/AC3F3D}

\bibitem[{Wordsworth \& Kreidberg(2022)}]{Wordsworth22_atmospheres}
Wordsworth, R., \& Kreidberg, L. 2022, Annual Review of Astronomy and
  Astrophysics, 60, 159, \dodoi{10.1146/annurev-astro-052920-125632}

\bibitem[{Worsley(2023)}]{Worsley23_irisesmfregrid}
Worsley, S. 2023, iris-esmf-regrid.
\newblock \url{https://github.com/SciTools-incubator/iris-esmf-regrid}

\bibitem[{Yang {et~al.}(2013)Yang, Cowan, \& Abbot}]{Yang13_stabilizing}
Yang, J., Cowan, N.~B., \& Abbot, D.~S. 2013, The Astrophysical Journal, 771,
  L45, \dodoi{10.1088/2041-8205/771/2/L45}

\bibitem[{Yang {et~al.}(2023)Yang, Zhang, Fu, Yan, Song, Wei, Liu, Ding, \&
  Tan}]{Yang23_cloud}
Yang, J., Zhang, Y., Fu, Z., {et~al.} 2023, Nature Astronomy, 1,
  \dodoi{10.1038/s41550-023-02015-8}

\bibitem[{Zhang {et~al.}(2017)Zhang, Tian, Wang, Dudhia, \&
  Chen}]{Zhang17_surface}
Zhang, X., Tian, F., Wang, Y., Dudhia, J., \& Chen, M. 2017, The Astrophysical
  Journal, 837, L27, \dodoi{10.3847/2041-8213/aa62fc}

\end{thebibliography}
\bibliographystyle{aasjournal}


\end{document}